\documentclass[10pt]{article}
\usepackage{amssymb}
\usepackage{amsmath}

\input{tcilatex}

\begin{document}

\title{Massless DKP field in Lyra manifold}
\author{R. Casana, C. A. M. de Melo and B. M. Pimentel \\
{\small Instituto de F\'{\i}sica Te\'orica, Universidade Estadual Paulista}
\vspace{-.1cm}\\
{\small Rua Pamplona 145, CEP 01405-900, S\~ao Paulo, SP, Brazil}}
\maketitle

\begin{abstract}
Massless scalar and vector fields are coupled to Lyra geometry by means of
Duffin-Kemmer-Petiau (DKP) theory. Using Schwinger Variational Principle,
equations of motion, conservation laws and gauge symmetry are implemented.
We find that the scalar field couples to the anholonomic part of the torsion
tensor, and the gauge symmetry of the electromagnetic field is not breaking
by the coupling with torsion.
\end{abstract}


\section{Introduction}

The existence and nature of the interaction of the gravitational torsion
with other fields is one of the most intriguing questions of the current
research in the classical theory of gravitation. The usual geometries
applied to model the gravitational interaction gives results that are not in
complete accordance with the physically expected. For instance, the Riemann
geometry completely ignores the presence of spin as a fundamental
characteristic of the matter field \cite{SabbataSpin}, but the quantum
experiences demonstrate that is in the same footing as the mass.
Riemann-Cartan geometry incorporate torsion as the geometric counterpart of
spin, in the same sense as curvature is manifested by the presence of
mass-energy. However, in the Einstein-Cartan theory \cite{sabbata} the
torsion is not a propagating quantity, remaining confined to the interior of
matter, which is not in complete accord with the idea of a \emph{dynamic}
space-time. Besides, the minimal coupling prescription leads to a coupling
among the electromagnetic field and torsion in such a way that the gauge
symmetry is broken in the Einstein-Cartan space-time. Several proposals
\emph{ad hoc} of construction of a propagating torsion are present in the
literature \cite{SabbataSpin, Saa}, but until now no one was a direct
consequence of the \emph{geometry} adopted to describe the space-time.

Here, we propose to use the Lyra geometry \cite{Lyra, Sen} to model the
gravitational content as a natural way to incorporate a dynamic torsion. In
the Lyra geometry, scale transformations are incorporated in the
structureless manifold, and all the differential geometry is constructed in
harmony with these scale transformations. In the next section a brief
introduction to Lyra geometry is given.

Our main objective here is to check if the hypothesis of the Lyra geometry
as a model of space-time is in good physical expectation about the torsion
behavior. To do it, we couple the Lyra manifold with massless bosonic fields
of spin $0$ and $1$ through the Duffin-Kemmer-Petiau theory \cite%
{HarishC,CQG}, which has as advantage an algebraically unified description
of both fields. The study of the coupling, equations of motion, conservation
laws and gauge transformations is performed by means of the Schwinger
Variational Principle \cite{QuanField, Sudarshan}, which is a powerful tool
to classical as well as quantum calculations.

Finally, some observations about the results found will be made.

\section{The Lyra geometry\label{LyraGeom}}


The Lyra manifold \cite{Lyra} is defined giving a tensor metric $g_{\mu \nu }
$ and a positive definite scalar function $\phi $ which we call the scale
function. In Lyra geometry one can change scale and coordinate system in an
independent way, to compose what is called a \emph{reference system}
transformation: let $M\subseteq \mathbb{R}^{N}$ and $U$ an open ball in $%
\mathbb{R}^{n}$, ($N\geq n$) and let $\chi :U\curvearrowright M$. The pair $%
\left( \chi ,U\right) $ defines a \emph{coordinate system}. Now, we define a
reference system by $\left( \chi ,U,\phi \right) $ where $\phi $ transforms
like
\begin{equation}
\bar{\phi}\left( \bar{x}\right) =\bar{\phi}\left( x\left( \bar{x}\right)
;\phi \left( x\left( \bar{x}\right) \right) \right) \quad ,\quad \frac{%
\partial \bar{\phi}}{\partial \phi }\not=0
\end{equation}%
under a reference system transformation.

In the Lyra's manifold, vectors transform as
\begin{equation}
\bar{A}^{\nu }=\frac{\bar{\phi}}{\phi }\frac{\partial \bar{x}^{\nu }}{%
\partial x^{\mu }}A^{\mu }
\end{equation}
In this geometry, the affine connection is
\begin{equation}
\tilde{\Gamma}^{\rho }{}_{\mu \nu }=\frac{1}{\phi }\mathring{\Gamma}^{\rho
}{}_{\mu \nu }+\frac{1}{\phi }\left[ \delta _{\,\mu }^{\rho }\partial _{\nu
}\ln \left( \frac{\phi }{\bar{\phi}}\right) -g_{\mu \nu }g^{\rho \sigma
}\partial _{\sigma }\ln \left( \frac{\phi }{\bar{\phi}}\right) \right]
\label{ConexLyra}
\end{equation}
whose transformation law is given by
\begin{equation}
\tilde{\Gamma}^{\rho }{}_{\mu \nu }=\frac{\bar{\phi}}{\phi }\bar{\Gamma}%
^{\sigma }{}_{\lambda \varepsilon }\frac{\partial x^{\rho }}{\partial \bar{x}%
^{\sigma }}\frac{\partial \bar{x}^{\lambda }}{\partial x^{\mu }}\frac{%
\partial \bar{x}^{\varepsilon }}{\partial x^{\nu }}+\frac{1}{\phi }\frac{%
\partial x^{\rho }}{\partial \bar{x}^{\sigma }}\frac{\partial ^{2}\bar{x}%
^{\sigma }}{\partial x^{\mu }\partial x^{\nu }}+\frac{1}{\phi }\delta
_{\,\nu }^{\rho }\frac{\partial }{\partial x^{\mu }}\ln \left( \frac{\bar{%
\phi}}{\phi }\right) \,.  \label{TransConexLyra}
\end{equation}

One can define the covariant derivative for a vector field as
\begin{equation*}
\nabla _{\mu }A^{\nu }={\frac{1}{\phi }}\partial _{\mu }A^{\nu }+\tilde{%
\Gamma}^{\nu }{}_{\mu \alpha }A^{\alpha }\,,\quad \nabla _{\mu }A_{\nu }={%
\frac{1}{\phi }}\,\partial _{\mu }A_{\nu }-\tilde{\Gamma}^{\alpha }{}_{\mu
\nu }A_{\alpha }\,.
\end{equation*}%
The metricity condition $\nabla _{\mu }\,g_{\nu \alpha }=0$ provided
\begin{equation}
\tilde{\Gamma}^{\rho }{}_{\mu \nu }=\frac{1}{\phi }\,\mathring{\Gamma}^{\rho
}{}_{\mu \nu }+\tilde{\Gamma}^{\rho }{}_{\left[ \mu \nu \right] }-\tilde{%
\Gamma}^{\sigma }{}_{\left[ \mu \lambda \right] }g_{\sigma \nu }g^{\rho
\lambda }-\tilde{\Gamma}^{\sigma }{}_{\left[ \nu \lambda \right] }g_{\sigma
\mu }g^{\rho \lambda }\,,
\end{equation}%
where
\begin{equation}
\mathring{\Gamma}^{\rho }{}_{\mu \nu }\equiv \frac{1}{2}g^{\rho \sigma
}\left( \partial _{\mu }g_{\nu \sigma }+\partial _{\nu }g_{\sigma \mu
}-\partial _{\sigma }g_{\mu \nu }\right)
\end{equation}%
is the analogous of the Levi-Civita connection and $\tilde{\Gamma}^{\rho
}{}_{\left[ \mu \nu \right] }$~is the antisymmetric part of the connection$\
$%
\begin{equation}
2\tilde{\Gamma}^{\rho }{}_{\left[ \mu \nu \right] }=\tilde{\Gamma}^{\rho
}{}_{\mu \nu }-\tilde{\Gamma}^{\rho }{}_{\nu \mu }\,=\frac{1}{\phi }\left(
\delta _{\mu }^{\rho }\partial _{\nu }-\delta _{\nu }^{\rho }\partial _{\mu
}\right) \ln \left( \frac{\phi }{\bar{\phi}}\right)
\end{equation}

The richness of the Lyra's geometry is demonstrated by\ the \emph{curvature}
\cite{Sen}
\begin{equation}
\tilde{R}^{\rho }{}_{\beta \alpha \sigma }\equiv \frac{1}{\phi ^{2}}\left(
\frac{\partial \left( \phi \tilde{\Gamma}^{\rho }{}_{\alpha \sigma }\right)
}{\partial x^{\beta }}-\frac{\partial \left( \phi \tilde{\Gamma}^{\rho
}{}_{\beta \sigma }\right) }{\partial x^{\alpha }}+\phi \tilde{\Gamma}^{\rho
}{}_{\beta \lambda }\phi \tilde{\Gamma}^{\lambda }{}_{\alpha \sigma }-\phi
\tilde{\Gamma}^{\rho }{}_{\alpha \lambda }\phi \tilde{\Gamma}^{\lambda
}{}_{\beta \sigma }\right)   \label{CurvLyra}
\end{equation}%
and the \emph{torsion}
\begin{equation}
\tilde{\tau}_{\mu \nu }{}^{\rho }=\tilde{\Gamma}^{\rho }{}_{\mu \nu }-\tilde{%
\Gamma}^{\rho }{}_{\nu \mu }-\frac{1}{\phi }\left( \delta _{\mu }^{\rho
}\partial _{\nu }-\delta _{\nu }^{\rho }\partial _{\mu }\right) \ln \phi
\label{torsion1}
\end{equation}%
where the second term is the anholonomic contribution, thus, we get
\begin{equation}
\tilde{\tau}_{\mu \nu }{}^{\rho }=-\frac{1}{\phi }\left( \delta _{\mu
}^{\rho }\partial _{\nu }-\delta _{\nu }^{\rho }\partial _{\mu }\right) \ln
\bar{\phi}  \label{torsion2}
\end{equation}%
which has intrinsic link with the scale functions and whose trace is given
by
\begin{equation}
\tilde{\tau}_{\mu \rho }{}^{\rho }\equiv \tilde{\tau}_{\mu }=\ \frac{3}{\phi
}\partial _{\mu }\ln \bar{\phi}\,.  \label{torsion-trace}
\end{equation}

In the next section we introduce the behavior of massless DKP field in the
Lyra geometry.

\section{The massless DKP field in Lyra manifold}

The massless DKP theory can not be obtained as a zero mass limit of the
massive DKP case, so we consider the Harish-Chandra Lagrangian density for
the massless DKP theory in the Minkowski space-time $\mathcal{M}^{4}$, given
by \cite{HarishC}
\begin{equation}
\mathcal{L_{M}}=i\bar{\psi}\gamma \beta ^{a}\partial _{a}\psi -i\partial _{a}%
\bar{\psi}\beta ^{a}\gamma \psi -\bar{\psi}\gamma \psi \;,  \label{minkowski}
\end{equation}%
where the $\beta ^{a}$ matrices satisfy the usual DKP algebra
\begin{equation}
\beta ^{a}\beta ^{b}\beta ^{c}+\beta ^{c}\beta ^{b}\beta ^{a}=\beta ^{a}\eta
^{bc}+\beta ^{c}\eta ^{ba}  \label{dkp-a1}
\end{equation}%
and $\gamma $ is a \textit{singular} matrix satisfying\footnote{%
We choose a representation in which ${\beta ^{0}}^{\dag }={\beta ^{0}}$, ${%
\beta ^{i}}^{\dag }=-{\beta ^{i}}$ and $\gamma ^{\dag }=\gamma $ \thinspace .%
}
\begin{equation}
\beta ^{a}\gamma +\gamma \beta ^{a}=\beta ^{a}\qquad ,\qquad \gamma
^{2}=\gamma \,.  \label{dkp-a2}
\end{equation}

From the above Lagrangian follows the massless DKP wave equation
\begin{equation}
i\beta ^{a}\partial _{a}\psi -\gamma \psi =0\;.  \label{eqm4}
\end{equation}

As it was known, the Minkowskian Lagrangian density (\ref{minkowski}) in its
massless spin 1 sector reproduce the electromagnetic or Maxwell theory with
its respective $U(1)$ local gauge symmetry.

To construct the covariant derivative of massless DKP field in Lyra
geometry, we follow the standard procedure of analyzing the behavior of the
field under local Lorentz transformations,
\begin{equation}
\psi \left( x\right) \rightarrow \psi ^{\prime }\left( x\right) =U\left(
x\right) \psi \left( x\right)  \label{LocalLorentz}
\end{equation}%
where $U$\ is a spin representation of Lorentz group characterizing the DKP
field. Now we define a \emph{spin connection} $S_{\mu }$ in a such way that
the object
\begin{equation}
\nabla _{\mu }\psi \equiv \frac{1}{\phi }\partial _{\mu }\psi +S_{\mu }{}\psi
\label{cov-fer}
\end{equation}%
transforms like a DKP field in (\ref{LocalLorentz}), thus, we set%
\begin{equation}
\nabla _{\mu }\psi \rightarrow \left( \nabla _{\mu }\psi \right) ^{\prime
}=U\left( x\right) \nabla _{\mu }\psi
\end{equation}%
and therefore $S$ transforms like
\begin{equation}
S_{\mu }^{\prime }=U\left( x\right) S_{\mu }U^{-1}\left( x\right) -\frac{1}{%
\phi }\left( \partial _{\mu }U\right) U^{-1}\left( x\right)
\label{TransfSpin}
\end{equation}

From the covariant derivative of the DKP field (\ref{cov-fer}) and
remembering that $\bar{\psi}\psi $ must be a scalar under the transformation
(\ref{LocalLorentz}), it follows that
\begin{equation}
\nabla _{\mu }\bar{\psi}=\frac{1}{\phi }\partial _{\mu }\bar{\psi}-\bar{\psi}%
S_{\mu }
\end{equation}

Then, we use the covariant derivative of the DKP current
\begin{equation*}
\nabla _{\mu }\left( \bar{\psi}\beta ^{\nu }\psi \right) =\frac{1}{\phi }%
\partial _{\mu }\left( \bar{\psi}\beta ^{\nu }\psi \right) +\Gamma ^{\nu
}{}_{\mu \lambda }\left( \bar{\psi}\beta ^{\lambda }\psi \right) =\left(
\nabla _{\mu }\bar{\psi}\right) \beta ^{\nu }\psi +\bar{\psi}\left( \nabla
_{\mu }\beta ^{\nu }\right) \psi +\bar{\psi}\beta ^{\nu }\left( \nabla _{\mu
}\psi \right)
\end{equation*}%
to get the following expression for the covariant derivative of $\beta ^{\nu
}$
\begin{equation}
\nabla _{\mu }\beta ^{\nu }=\frac{1}{\phi }\partial _{\mu }\beta ^{\nu
}+\Gamma ^{\nu }{}_{\mu \lambda }\beta ^{\lambda }+S_{\mu }\beta ^{\nu
}-\beta ^{\nu }S_{\mu }  \label{cov-beta}
\end{equation}

To solve the equation above we introduce the tetrad field $e^{\mu }{}_{a}$
and its inverse $e_{\mu }{}^{a}$ related to the space-time metric by the
following equations
\begin{align}
g^{\mu \nu }(x)& =\eta ^{ab}\,e^{\mu }{}_{a}(x)e^{\nu }{}_{b}(x)\,,  \notag
\\
g_{\mu \nu }(x)& =\eta _{ab}e_{\mu }{}^{a}(x)e_{\nu }{}^{b}(x)\,, \\
e_{\mu }{}^{a}(x)& =g_{\mu \nu }(x)\eta ^{ab}e^{\nu }{}_{b}(x)  \notag
\end{align}

Such that the metricity condition, $\nabla _{\mu }g_{\alpha \beta }=0$, can
be expressed as
\begin{equation}
\nabla _{\mu }e^{\nu }{}_{a}=\frac{1}{\phi }\partial _{\mu }e^{\nu
}{}_{a}+\Gamma ^{\nu }{}_{\mu \lambda }e^{\lambda }{}_{a}+{}\omega _{\mu
a}{}^{b}e^{\nu }{}_{b}=0  \label{tetrad-metric}
\end{equation}%
where $\omega _{\mu ab}$ is the spin connection coefficients. Expressing $%
\beta ^{\nu }$ in terms of the tetrad fields, $\beta ^{\nu }=e^{\nu
}{}_{a}\beta ^{a}$, in the equation (\ref{cov-beta}) we get
\begin{equation}
\omega _{\mu a}{}^{b}e^{\nu }{}_{b}\beta ^{a}=S_{\mu }\beta ^{\nu }-\beta
^{\nu }S_{\mu }=\left[ S_{\mu },\beta ^{\nu }\right]
\end{equation}%
from which we found that
\begin{equation}
S_{\mu }=\frac{1}{2}\omega _{\mu ab}S^{ab}
\end{equation}%
with
\begin{equation}
S^{ab}=\left[ \beta ^{a},\beta ^{b}\right] .
\end{equation}

With a covariant derivative of the DKP\ field well-defined we can consider
the Lagrangian density (\ref{minkowski}) of the massless DKP field minimally
coupled \cite{misner,hehl} to the Lyra manifold, thus, the action reads
\begin{equation}
S=\int_{\Omega }d^{4}x~\phi ^{4}e\left( \frac{{}}{{}}\!i\,\bar{\psi}\gamma
e_{\;\,a}^{\mu }\beta ^{a}\nabla _{\mu }\psi -i\nabla _{\mu }\bar{\psi}\beta
^{a}e_{\;\,a}^{\mu }\gamma \psi -\bar{\psi}\gamma \psi \right) \;.
\label{lag-cv}
\end{equation}%
where $\nabla _{\mu }$ is the Lyra covariant derivative of DKP field defined
above.

\section{Dynamics of the massless DKP field coupled to the Lyra manifold}

In following we use a classical version of the Schwinger Action Principle
such as it was treated in the context of Classical Mechanics by Sudarshan
and Mukunda \cite{Sudarshan}. The Schwinger Action Principle is the most
general version of the usual variational principles. It was proposed
originally at the scope of the Quantum Field Theory \cite{QuanField}, but
its application goes beyond this area. In the classical context, the basic
statement of the Schwinger Principle is%
\begin{equation}
\delta S=\delta \int_{\Omega }dx~e\phi ^{4}\mathcal{L}=\int_{\partial \Omega
}d\sigma _{\mu }G^{\mu }
\end{equation}%
where $S$ is the classical actions and the $G^{\mu }$'s are the generators
of the canonical transformations. The Schwinger Principle can be employed,
choosing suitable variations in each case, to obtain commutation relations
in the quantum context or canonical transformations in the classical one, as
well as equations of motion or still perturbative expansions.

Here, we will apply the Action Principle to derive equations of motion of
the massless DKP field in an external Lyra background and its conservations
laws associated with translations and rotations in such space.

Making the total variation of the action integral (\ref{lag-cv}) we get%
\begin{eqnarray}
\delta S &=&\int_{\Omega }dx~e\phi ^{4}\left[ 4\mathcal{L}-\frac{i}{\phi }%
\bar{\psi}\gamma \beta ^{\mu }\partial _{\mu }\psi +\frac{i}{\phi }\partial
_{\mu }\bar{\psi}\beta ^{\mu }\gamma \psi \right] \left( \frac{\delta \phi }{%
\phi }\right) +\int_{\Omega }dx~\phi ^{4}e\left( \frac{\delta e}{e}\right)
\mathcal{L}+  \label{action-variation} \\
&&+\int_{\Omega }dx~e\phi ^{4}\left[ \frac{{}}{{}}i\bar{\psi}\gamma \left(
\delta \beta ^{\mu }\right) \nabla _{\mu }\psi -i\nabla _{\mu }\bar{\psi}%
\left( \delta \beta ^{\mu }\right) \gamma \psi \right] ~+\int_{\Omega
}dx~e\phi ^{4}\left[ \frac{{}}{{}}i\bar{\psi}\gamma \beta ^{\mu }\left(
\delta S_{\mu }\right) \psi +i\bar{\psi}\left( \delta S_{\mu }\right) \beta
^{\mu }\gamma \psi \right] +  \notag \\
&&+\int_{\Omega }dx~e\phi ^{4}\left[ \left( \delta \bar{\psi}\right) \left(
\frac{{}}{{}}i\gamma \beta ^{\mu }\nabla _{\mu }\psi -\gamma \psi +iS_{\mu
}\beta ^{\mu }\gamma \psi \right) -\frac{i}{\phi }\left( \delta \partial
_{\mu }\bar{\psi}\right) \beta ^{\mu }\gamma \psi \right] +  \notag \\
&&-\int_{\Omega }dx~e\phi ^{4}\left[ \left( \frac{{}}{{}}i\nabla _{\mu }\bar{%
\psi}\beta ^{\mu }\gamma +\bar{\psi}\gamma -i\bar{\psi}\gamma \beta ^{\mu
}S_{\mu }\right) \delta \psi -\frac{i}{\phi }\bar{\psi}\gamma \beta ^{\mu
}\left( \delta \partial _{\mu }\psi \right) \right]  \notag
\end{eqnarray}%
where $\mathcal{L}$ is the Lagrangian density in (\ref{lag-cv}). Choosing
different specializations of the variations, one can easily obtain the
equations of motion and the energy-momentum and spin density tensors.

\subsection{Equations of motion}

To begin, we choose to make functional variations only in the massless DKP
field thus we set $\delta \phi =\delta e^{\mu }{}_{b}=\delta \omega _{\mu
ab}=0$ and considering $\left[ \delta ,\partial _{\mu }\right] =0$, from (%
\ref{action-variation}) we have, after an integration by parts,
\begin{eqnarray}
\delta S &=&\int_{\partial \Omega }d\sigma _{\mu }~e\phi ^{3}i\left[ \frac{{}%
}{{}}\bar{\psi}\gamma \beta ^{\mu }\left( \delta \psi \right) -\left( \delta
\bar{\psi}\right) \beta ^{\mu }\gamma \psi \right] +  \label{lyraEQ-32} \\
&&+\int_{\Omega }dx~e\phi ^{4}\left( \delta \bar{\psi}\right) \left[ \frac{{}%
}{{}}i\beta ^{\mu }\nabla _{\mu }\psi +i\tilde{\tau}_{\mu }\beta ^{\mu
}\gamma \psi -\gamma \psi \right] -\int_{\Omega }dx~e\phi ^{4}\left[ \frac{{}%
}{{}}i\nabla _{\mu }\bar{\psi}\beta ^{\mu }+i\tilde{\tau}_{\mu }\bar{\psi}%
\gamma \beta ^{\mu }+\bar{\psi}\gamma \right] \delta \psi  \notag
\end{eqnarray}
where $\tilde{\tau}_{\mu }$ is the trace torsion and, it is given by
\begin{equation}
\tilde{\tau}_{\mu }=\tilde{\tau}_{\mu \rho }{}^{\rho }=\frac{3}{\phi }%
\partial _{\mu }\ln \bar{\phi}~\ .  \label{traza-torsion}
\end{equation}

Following the action principle we get the generator of the variations of the
massless DKP field%
\begin{equation}
G_{\delta \psi }=\int_{\partial \Omega }d\sigma _{\mu }~e\phi ^{3}i\left[
\frac{{}}{{}}\bar{\psi}\gamma \beta ^{\mu }\left( \delta \psi \right)
-\left( \delta \bar{\psi}\right) \beta ^{\mu }\gamma \psi \right]
\label{lyraEQ-33}
\end{equation}%
and its equations of motion in the Lyra manifold,%
\begin{eqnarray}
i\beta ^{\mu }\left( \nabla _{\mu }+\tilde{\tau}_{\mu }\gamma \right) \psi
-\gamma \psi &=&0  \notag \\
&&  \label{lyraEQ-34} \\
i\nabla _{\mu }\bar{\psi}\beta ^{\mu }+i\tilde{\tau}_{\mu }\bar{\psi}\gamma
\beta ^{\mu }+\bar{\psi}\gamma &=&0  \notag
\end{eqnarray}

\subsection{Local gauge symmetry}

We set $\delta \phi =\delta e^{\mu }{}_{b}=\delta \omega _{\mu ab}=0$ in (%
\ref{action-variation}) thus we get
\begin{eqnarray}
\delta S &=&\int_{\Omega }dx~e\phi ^{4}\left[ ~\left( \delta \bar{\psi}%
\right) \left( \frac{{}}{{}}i\gamma \beta ^{\mu }\nabla _{\mu }\psi -\gamma
\psi +iS_{\mu }\beta ^{\mu }\gamma \psi \right) -\frac{i}{\phi }\left(
\partial _{\mu }\delta \bar{\psi}\right) \beta ^{\mu }\gamma \psi \right] +
\label{lyraEQ-35} \\
&&-\int_{\Omega }dx~e\phi ^{4}\left[ \left( \frac{{}}{{}}i\nabla _{\mu }\bar{%
\psi}\beta ^{\mu }\gamma +\bar{\psi}\gamma -i\bar{\psi}\gamma \beta ^{\mu
}S_{\mu }\right) \left( \delta \psi \right) -\frac{i}{\phi }\bar{\psi}\gamma
\beta ^{\mu }\left( \partial _{\mu }\delta \psi \right) \right]  \notag
\end{eqnarray}%
and choosing the local variation of the fields as being%
\begin{equation}
\delta \psi =\left( 1-\gamma \right) \Phi ~\ \ \ \ \ ~~\ ,~\ \ \ ~~\ \ \
\delta \bar{\psi}=\bar{\Phi}\left( 1-\gamma \right)  \label{lyraEQ-36}
\end{equation}%
then the variation (\ref{lyraEQ-35})\ reads as
\begin{equation}
\delta S=\int_{\Omega }dxe\phi ^{4}\left( \bar{\psi}i\beta ^{\mu }\left[
\nabla _{\mu }\left( 1-\gamma \right) \Phi \right] -i\left[ \nabla _{\mu }%
\bar{\Phi}\left( 1-\gamma \right) \right] \beta ^{\mu }\psi \right)
\end{equation}

If we impose that the variations (\ref{lyraEQ-36}) gives rise to a symmetry,
$\delta S=0$, then the fields $\Phi $ and $\bar{\Phi}$ must satisfy%
\begin{eqnarray}
i\beta ^{\mu }\nabla _{\mu }\left( 1-\gamma \right) \Phi &=&0  \notag \\
&&  \label{lyraEQ-37} \\
i\nabla _{\mu }\bar{\Phi}\left( 1-\gamma \right) \beta ^{\mu } &=&0  \notag
\end{eqnarray}%
Under such conditions the local transformation
\begin{eqnarray}
\psi &\rightarrow &\psi +(1-\gamma )\Phi  \notag \\
&&  \label{lyraEQ-38} \\
\bar{\psi} &\rightarrow &\bar{\psi}+\bar{\Phi}(1-\gamma )  \notag
\end{eqnarray}%
is a local gauge symmetry of the action (\ref{lag-cv}) and the equations of
motion (\ref{lyraEQ-34}).

\subsection{Energy-momentum tensor and spin tensor density}

Now, we vary only the background manifold and we assume that $\delta \omega
_{\mu ab}$\ and $\delta e^{\mu }{}_{a}$ are independent variations, the
general variation (\ref{action-variation}) reads
\begin{equation}
\delta S=\int_{\Omega }dx~e\phi ^{4}\left\{ \left[ i\left( \frac{{}}{{}}\bar{%
\psi}\gamma \beta ^{a}\nabla _{\mu }\psi -\nabla _{\mu }\bar{\psi}\beta
^{a}\gamma \psi \right) -e_{\mu }{}^{a}\mathcal{L}\right] \delta e^{\mu
}{}_{a}~+i\left( \frac{{}}{{}}\bar{\psi}\gamma \beta ^{\mu }S^{ab}\psi +\bar{%
\psi}S^{ab}\beta ^{\mu }\gamma \psi \right) \frac{1}{2}\delta \omega _{\mu
ab}\right\} .  \label{VarAcBack}
\end{equation}%
where we have used $\delta e=-ee_{\mu }{}^{a}\delta e^{\mu }{}_{a}$.

First, holding only the variations in the tetrad field, $\delta \omega _{\mu
ab}=0$, we found for the variation of the action
\begin{equation}
\delta S=\int_{\Omega }dx~e\phi ^{4}\left[ \frac{{}}{{}}i\bar{\psi}\gamma
\beta ^{a}\nabla _{\mu }\psi -i\nabla _{\mu }\bar{\psi}\beta ^{a}\gamma \psi
-e_{\mu }{}^{a}\mathcal{L}\right] \delta e^{\mu }{}_{a}\,.
\end{equation}

Defining the energy-momentum density tensor as
\begin{equation}
T_{\mu }{}^{a}\equiv \frac{1}{\phi ^{4}e}\frac{\delta S}{\delta e^{\mu
}{}_{a}}=i\bar{\psi}\gamma \beta ^{a}\nabla _{\mu }\psi -i\nabla _{\mu }\bar{%
\psi}\beta ^{a}\gamma \psi -e_{\mu }{}^{a}\mathcal{L}
\label{energy-momentum}
\end{equation}%
which can be written in coordinates as $T_{\mu }{}^{\nu }\equiv e^{\nu
}{}_{a}T_{\mu }{}^{a}$. \

Now, making functional variations only in the components of the spin
connection, $\delta e^{\mu }{}_{a}=0$, we found for the action variation
\begin{equation}
\delta S=\int_{\Omega }dx~e\phi ^{4}~\frac{1}{2}\left( \delta \omega _{\mu
ab}\right) i\bar{\psi}\left( \frac{{}}{{}}\!\gamma \beta ^{\mu
}S^{ab}+S^{ab}\beta ^{\mu }\gamma \right) \psi ,
\end{equation}%
we define the spin density tensor as being
\begin{equation}
S^{\mu ab}\equiv \frac{2}{\phi ^{4}e}\frac{\delta S}{\delta \omega _{\mu ab}}%
=i\bar{\psi}\left( \frac{{}}{{}}\!\gamma \beta ^{\mu }S^{ab}+S^{ab}\beta
^{\mu }\gamma \right) \psi \,.  \label{spin-tensor}
\end{equation}

\subsection{Functional Scale Variations and the Trace Relation}

Under a pure infinitesimal variation $\delta \phi $ of the scale function
the action variation (\ref{action-variation}) is expressed as%
\begin{equation}
\delta S=-\int_{\Omega }dx~e\phi ^{4}\left( T_{\mu }{}^{a}e^{\mu }{}_{a}-%
\frac{1}{2}S^{\mu ab}\omega _{\mu ab}\right) \frac{\delta \phi }{\phi }\,,
\end{equation}%
where we have used the definition of the energy-momentum and spin density
tensors. From it we obtain the following algebraic property satisfied by the
energy-momentum tensor trace
\begin{equation}
T_{\mu }{}^{a}e^{\mu }{}_{a}-\frac{1}{2}S^{\mu ab}\omega _{\mu ab}=0\,,
\label{trace-identity}
\end{equation}
which is the so called trace relation. Such identity can be used to
constraint the form of the connection $\omega $ in a given content of matter.

\subsection{Conservation laws}

As another application of the Schwinger Action Principle, let us to derive
the conservation laws associated to local Lorentz transformations and
infinitesimal general coordinate transformations.

\subsubsection{Local Lorentz transformations}

Under \emph{local Lorentz transformations}, the functional variations of the
tetrad and the spin connection are given by%
\begin{equation}
\delta e^{\mu }{}_{a}=\delta \varepsilon _{a}{}^{b}~e^{\mu }{}_{b}\,
\label{var-tetra}
\end{equation}%
\begin{equation}
\delta \omega _{\mu ab}=\omega _{\mu }{}^{c}{}_{b}\,\delta \varepsilon
_{ac}-\omega _{\mu {a}}{}^{c}\,\delta \varepsilon _{cb}-\frac{1}{\phi }%
\partial _{\mu }\delta \varepsilon _{ab}  \label{var-spin}
\end{equation}%
with $\delta \varepsilon _{ab}=-\delta \varepsilon _{ba}$, where the first
variation express the vectorial character of the tetrad in Minkowski
space-time, and the second one comes from (\ref{TransfSpin}) with
\begin{equation}
U=1+\frac{1}{2}\delta \varepsilon _{ab}\Sigma ^{ab}.
\end{equation}

The general expression (\ref{VarAcBack}) can be written using the
definitions of energy-momentum and spin tensors as being%
\begin{equation}
\delta S=\int_{\Omega }dx~e\phi ^{4}\left( T_{\mu }{}^{a}\delta e^{\mu
}{}_{a}+S^{\mu ab}\frac{1}{2}\delta \omega _{\mu ab}\right) .  \label{tgc-1}
\end{equation}%
Substituting variations (\ref{var-tetra}) and (\ref{var-spin}) and after
some integration by parts and algebraic manipulations we get
\begin{equation}
\delta S=-\int_{\partial \Omega }d\sigma _{\mu }\left( \phi ^{3}eS^{\mu ab}%
\frac{1}{2}\delta \varepsilon _{ab}\right) +\int_{\Omega }dx\phi ^{4}e\left(
\frac{{}}{{}}\!\nabla _{\mu }S^{\mu ab}+\tilde{\tau}_{\mu }S^{\mu
ab}-T^{ab}+T^{ba}\right) \frac{1}{2}\delta \varepsilon _{ab}
\end{equation}%
where $\tilde{\tau}_{\mu }$ is the trace torsion defined in (\ref%
{torsion-trace}). Thus, from the Action Principle, we get $G_{\delta
\varepsilon }$ the generator of infinitesimal changes in the DKP field under
local Lorentz transformations
\begin{equation}
G_{\delta \varepsilon }=-\int_{\partial \Omega }d\sigma _{\mu }\left( \phi
^{3}eS^{\mu ab}\frac{1}{2}\delta \varepsilon _{ab}\right)  \label{spin-gen}
\end{equation}%
and the conservation law for the spinning content of the theory
\begin{equation}
\nabla _{\mu }S^{\mu ab}+\tilde{\tau}_{\mu }S^{\mu ab}=T^{ab}-T^{ba}
\label{ConservSpin}
\end{equation}%
The most important aspect here is the coupling among the spin density tensor
and the torsion.

\subsection{General coordinate transformation}

Now let us to perform a \emph{general coordinate transformation} in the
action. From the Lyra transformation rule for vectors and the infinitesimal
displacement $\bar{x}^{\mu }=x^{\mu }+\delta x^{\mu }$, we have that the
tetrad field transforms as
\begin{equation}
\bar{e}^{\mu }{}_{a}\left( \bar{x}\right) =\frac{\bar{\phi}\left( \bar{x}%
\right) }{\phi \left( x\right) }\frac{\partial \bar{x}^{\mu }}{\partial
x^{\nu }}e^{\nu }{}_{a}\left( x\right) \,,
\end{equation}%
with its form variation $\delta e^{\mu }{}_{a}\left( x\right) \equiv \bar{e}%
^{\mu }{}_{a}(x)-e^{\mu }{}_{a}\left( x\right) $ given by%
\begin{equation}
\delta e^{\mu }{}_{a}\left( x\right) =\frac{\bar{\phi}\left( x\right) }{\phi
\left( x\right) }\left[ e^{\mu }{}_{a}\left( x\right) +e^{\nu
}{}_{a\,}\left( x\right) \partial _{\nu }\delta x^{\mu }-\delta x^{\nu
}\partial _{\nu }e^{\mu }{}_{a}\left( x\right) +e^{\mu }{}_{a}\left(
x\right) \delta x^{\nu }\partial _{\nu }\ln \phi \left( x\right) \right]
-e^{\mu }{}_{a}\left( x\right) \,.  \label{gct-tetrad}
\end{equation}%
While the spin connection transforms as
\begin{equation}
\bar{\omega}_{\mu ab}^{\,}\left( \bar{x}\right) =\frac{\phi \left( x\right)
}{\bar{\phi}\left( \bar{x}\right) }\frac{\partial x^{\nu }}{\partial \bar{x}%
^{\mu }}\omega _{\nu ab}\left( x\right) \,,
\end{equation}%
with its form variation $\delta \omega _{\mu ab}\left( x\right) \equiv \bar{%
\omega}_{\mu ab}\left( x\right) -\omega _{\mu ab}\left( x\right) $ given as
\begin{equation}
\delta \omega _{\mu ab}\left( x\right) =\frac{\phi \left( x\right) }{\bar{%
\phi}\left( x\right) }\left[ \omega _{\mu ab}\left( x\right) -\omega _{\nu
ab}\left( x\right) \partial _{\mu }\delta x^{\nu }-\delta x^{\nu }\partial
_{\nu }\omega _{\mu ab}\left( x\right) -\omega _{\mu ab}\left( x\right)
\delta x^{\nu }\partial _{\nu }\ln \phi \left( x\right) \right] -\omega
_{\mu ab}\left( x\right) \,.  \label{gct-spinc}
\end{equation}

Substituting the expressions (\ref{gct-tetrad}) and (\ref{gct-spinc}) in ( %
\ref{tgc-1}),%
\begin{eqnarray}
\delta S &=&\int_{\partial \Omega }d\sigma _{\mu }~e\phi ^{4}\left( \frac{%
\bar{\phi}}{\phi }T_{\nu }{}^{\mu }-\frac{\phi }{\bar{\phi}}S^{\mu ab}\frac{1%
}{2}\omega _{\nu ab}\right) \delta x^{\nu }+\int_{\Omega }dx~e\phi
^{4}\left( \frac{\bar{\phi}}{\phi }-1\right) \left( T_{\mu }{}^{a}e^{\mu
}{}_{a}-\frac{\phi }{\bar{\phi}}\frac{1}{2}\omega _{\mu ab}S^{\mu ab}\right)
+  \notag \\
&&-\int_{\Omega }dx~e\phi ^{5}\left[ \frac{\bar{\phi}}{\phi }\left( \nabla
_{\mu }T_{\nu }{}^{\mu }+\tilde{\tau}_{\mu \nu }{}^{\lambda }T_{\lambda
}{}^{\mu }+\tilde{\tau}_{\mu }T_{\nu }{}^{\mu }\right) +\frac{\bar{\phi}}{%
\phi ^{2}}\left( \partial _{\mu }\ln \frac{\bar{\phi}}{\phi }\right) T_{\nu
}{}^{\mu }+\right.  \label{coordenada} \\
&&\ \ \ \ \ \ \ \ \ \ \ \ \ \ \ \ \ \ \ \left. ~~+\left( \frac{\bar{\phi}}{%
\phi }-\frac{\phi }{\bar{\phi}}\right) \omega _{\nu ab}T^{ab}+\frac{1}{2}%
\frac{1}{\bar{\phi}}\left( \partial _{\mu }\ln \frac{\bar{\phi}}{\phi }%
\right) S^{\mu ab}\omega _{\nu ab}+\frac{1}{2}\frac{\phi }{\bar{\phi}}S^{\mu
ab}R_{\nu \mu ab}\right] \delta x^{\nu }  \notag
\end{eqnarray}

According with the Schwinger Principle, we get the generator $G_{\delta x}$
\begin{equation}
G_{\delta x}=\int_{\partial \Omega }d\sigma _{\mu }~e\phi ^{4}\left( \frac{%
\bar{\phi}}{\phi }~T_{\nu }{}^{\mu }-\frac{\phi }{\bar{\phi}}S^{\mu ab}\frac{%
1}{2}\omega _{\nu ab}\right) \delta x^{\nu }~  \label{1ra-1}
\end{equation}
that establishes the form of the variations under infinitesimal coordinate
transformations.

Next, from the third integral in (\ref{coordenada}) and due to the
invariance of the action under general coordinate transformations we obtain
the conservation law of energy-momentum.
\begin{eqnarray}
\frac{\bar{\phi}}{\phi }\left( \nabla _{\mu }T_{\nu }{}^{\mu }+\tilde{\tau}%
_{\mu \nu }{}^{\lambda }T_{\lambda }{}^{\mu }+\tilde{\tau}_{\mu }T_{\nu
}{}^{\mu }\right) &\!\!\!+&\!\!\!\frac{1}{2}\frac{\phi }{\bar{\phi}}S^{\mu
ab}R_{\nu \mu ab}+  \notag  \label{3ra-3} \\
&& \\
+\left( \frac{\bar{\phi}}{\phi }-\frac{\phi }{\bar{\phi}}\right) \omega
_{\nu ab}T^{ab} &\!\!\!+&\!\!\!\frac{1}{\bar{\phi}}\left( \partial _{\mu
}\ln \frac{\bar{\phi}}{\phi }\right) \left[ \frac{\bar{\phi}^{2}}{\phi ^{2}}%
T_{\nu }{}^{\mu }+\frac{1}{2}S^{\mu ab}\omega _{\nu ab}\right] \;=\;0\,.
\notag
\end{eqnarray}

By imposing the invariance of the action under general coordinate
transformations, the second integral in (\ref{coordenada}) allows to get a
new relation, we named it as the trace symmetry. Such identity for $\,\bar{%
\phi}\neq \phi \,$ can be written as
\begin{equation}
T_{\mu }{}^{a}e^{\mu }{}_{a}-\frac{\phi }{\bar{\phi}}\frac{1}{2}S^{\mu
ab}\omega _{\mu ab}=0\,,  \label{2da-3}
\end{equation}%
it can be considered as the generalization of the trace relation shown in (%
\ref{trace-identity}) and, which relates the different scale functions with
the geometry and the field content. For $\bar{\phi}\equiv \phi $ it reduces
to (\ref{trace-identity}).


\section{The spin content}

The massless DKP theory describes in a algebraically unified way the spin $0$
and $1$ fields. However, scalar and vector fields have different particular
behaviours on couplings and gauge transformations. To see how this can be
implemented in the equations above, we use the projectors of Umezawa \cite%
{Umezawa, Lunardi1} to select these distinct spin sectors.

\subsection{Spin $0$ sector}

The spin $0$ projectors $P$ and $P^{\mu }$($=e^{\mu }{}_{a}P^{a}$) are such
that $P\psi $ and $P^{\mu }\psi $ transform, respectively, as a scalar and a
vector in the Lyra space-time. Then, first we apply the scalar projector $P$
in the equation of motion (\ref{lyraEQ-34}) and we get
\begin{equation*}
P\,\ \ \rightarrow \,\ \ \ \ i\nabla _{\mu }\left( P^{\mu }\psi \right) +i%
\tilde{\tau}_{\mu }\left( P^{\mu }\gamma \psi \right) -P\gamma \psi =0
\end{equation*}%
we multiply it by $\left( 1-\gamma \right) $ and get
\begin{equation}
i\left( \nabla _{\mu }+\tilde{\tau}_{\mu }\right) \left( P^{\mu }\gamma \psi
\right) =0.  \label{eq18}
\end{equation}%
Next, by applying the vector projector $P^{\mu }$ we obtain
\begin{equation*}
P^{\mu }\,\ \ \rightarrow \,\ \ \ \ i\nabla ^{\mu }\left( P\psi \right) +i%
\tilde{\tau}_{\mu }\left( P\gamma \psi \right) -P^{\mu }\gamma \psi =0
\end{equation*}%
and remembering that $P\gamma =\gamma P$, we obtain
\begin{equation}
P^{\mu }\gamma \psi =i\left( \nabla ^{\mu }+\tilde{\tau}^{\mu }\gamma
\right) \left( P\psi \right) .  \label{eq19}
\end{equation}%
By mixing the equations (\ref{eq18}) and (\ref{eq19}), we find the equation
of motion for the\ scalar field $P\psi $ in the Lyra space-time
\begin{equation}
\left( \nabla _{\mu }+\tilde{\tau}_{\mu }\right) (\nabla ^{\mu }+\tilde{\tau}%
_{\mu }\gamma )P\psi =0  \label{eq20}
\end{equation}

We use a specific representation of the DKP algebra in which the singular $%
\gamma $ matrix is
\begin{equation*}
\gamma =\mbox{diag}(\lambda ,1-\lambda ,1-\lambda ,1-\lambda ,1-\lambda ),
\end{equation*}%
such that the condition $\gamma ^{2}-\gamma =0$ implies that $\lambda $ $\in
\left\{ 0,1\right\} $. \ In Minkowski space--time, the value $\lambda =0$
reproduces the massless Klein-Gordon-Fock field \cite{CQG}. Thus, we
restrict our attention for the $\lambda =0$ case only. The scalar sector of
massless DKP theory can be explicitly worked out by using the five
dimensional representation of massless DKP algebra. Thus, the field $\psi $
is given by a 5-component column vector
\begin{equation}
\psi =\left(
\begin{array}{c}
\varphi ,\psi ^{0},\psi ^{1},\psi ^{2},\psi ^{3}%
\end{array}%
\right) ^{T}\,,
\end{equation}%
where $\varphi $ and $\psi ^{a}$ ($a=0,1,2,3$) behave respectively as a
scalar and a 4-vector under \textit{Lorentz} transformations on the
Minkowski space. And the other projections are
\begin{equation*}
P\psi =\left(
\begin{array}{c}
\varphi ,0,0,0,0%
\end{array}%
\right) ^{T}\quad ,\qquad \,P^{\mu }\psi =\left(
\begin{array}{c}
\psi ^{\mu },0,0,0,0%
\end{array}%
\right) ^{T}~\ \ ,
\end{equation*}%
\begin{equation*}
\gamma \psi =\left(
\begin{array}{c}
0,\psi ^{0},\psi ^{1},\psi ^{2},\psi ^{3}%
\end{array}%
\right) ^{T}\quad ,\qquad P\gamma \psi =0\,\quad ,\qquad P^{\mu }\gamma \psi
=\left(
\begin{array}{c}
\psi ^{\mu },0,0,0,0%
\end{array}%
\right) ^{T}~\ \ .
\end{equation*}%
Expressing the massless DKP action (\ref{lag-cv}) in this representation we
obtain
\begin{equation}
S_{0}=\int dx\phi ^{4}e\left[ i\psi ^{\ast \mu \,}\nabla _{\mu }\varphi
-i\psi ^{\mu }\nabla _{\mu }\varphi ^{\ast }-\psi ^{\ast a}\psi _{a}\right]
\,.  \label{eq21}
\end{equation}%
and from the equation (\ref{eq19}) we get the vectorial component of the DKP
field
\begin{equation}
\psi ^{\mu }=i\nabla ^{\mu }\varphi   \label{eq22}
\end{equation}%
and together with the equation (\ref{eq18}) \ we get the equation of motion
for the scalar field $\varphi $
\begin{equation}
\left( \nabla _{\mu }+\tilde{\tau}_{\mu }\right) \nabla ^{\mu }\varphi =0
\end{equation}%
or
\begin{equation}
\frac{1}{\phi }\left( \partial _{\mu }\partial ^{\mu }\varphi +\mathring{%
\Gamma}^{\mu }{}_{\mu \alpha }\partial ^{\alpha }\varphi \right) +\frac{2}{%
\phi }\left( \partial ^{\mu }\varphi \right) \partial _{\mu }\ln \left( \phi
\right) =0
\end{equation}%
Thus, contrary to what happens in the Riemann-Cartan case \cite{CQG}, we
conclude that \emph{the spin }$0$\emph{\ sector of the massless DKP field
does interact with the anholonomic component of the Lyra torsion}.

Finally, in this representation the action for the spin $0$ sector of the
massless DKP field in Lyra space-time is getting by substituting equation (%
\ref{eq22}) into the action (\ref{eq21}) \ which reduces to the usual one
obtained from the Minkowski Klein-Gordon-Fock lagrangian density minimally
coupled to Lyra space-time $\mathcal{\ }$
\begin{equation}
S_{0}=\int dx\,\phi ^{4}e\,\nabla _{\mu }\varphi ^{\ast }\nabla ^{\mu
}\varphi \,.  \label{eq27}
\end{equation}

The gauge transformation (\ref{lyraEQ-38}) reads as
\begin{equation*}
\varphi ^{\prime }=\varphi +\varphi _{_{\Phi }}\quad ,\quad \psi ^{\prime
}{}^{\mu }=\psi ^{\mu }\,,
\end{equation*}
while condition (\ref{lyraEQ-37}) becomes ${\nabla }_{\mu }\varphi _{_{\Phi
}}=\frac{1}{\phi }\partial _{\mu }\varphi _{_{\Phi }}=0$, i.e., $\varphi
_{_{\Phi }}$ must be a constant.


\subsection{Spin $1$ sector}


The spin 1 projectors $R^{\mu }$($=e^{\mu }{}_{a}R^{a}$ ) and $R^{\mu \nu }$(%
$=e^{\mu }{}_{a}e^{\nu }{}_{b}R^{ab}$) are such that $R^{\mu }\psi $ and $%
R^{\mu \nu }\psi $ transform respectively as a vector and a second rank
tensor under general coordinate transformations. Thus, using the projector $%
R^{\mu }$ in the equation of motion (\ref{lyraEQ-34}) we have
\begin{equation*}
\ i\nabla _{\nu }\left( R^{\mu \nu }\psi \right) +i\tilde{\tau}_{\nu }\left(
R^{\mu \nu }\gamma \psi \right) -R^{\mu }\gamma \psi =0
\end{equation*}%
multiplying by $\left( 1-\gamma \right) $ we get
\begin{equation}
i\left( \nabla _{\nu }+\tilde{\tau}_{\nu }\right) \left( R^{\mu \nu }\gamma
\psi \right) =0  \label{eq23}
\end{equation}%
where we have used $R^{\mu }\gamma =\gamma R^{\mu }$. \ Next, applying the
projector $R^{\mu \nu }$ we obtain
\begin{equation*}
i\nabla _{\rho }\left( R^{\mu \nu }\beta ^{\rho }\psi \right) +i\tilde{\tau}%
_{\rho }\left( R^{\mu \nu }\beta ^{\rho }\gamma \psi \right) -R^{\mu \nu
}\gamma \psi =0,
\end{equation*}%
using the properties $R^{\mu \nu }\beta ^{\rho }=R^{\mu }g^{\nu \rho
}-R^{\nu }g^{\mu \rho }$ and $R^{\mu \nu }\gamma =(1-\gamma )R^{\mu \nu }$,
we get
\begin{equation}
R^{\mu \nu }\gamma \psi =i\left( \nabla _{\rho }+\tilde{\tau}_{\rho }\gamma
\right) \left[ g^{\rho \nu }\left( R^{\mu }\psi \right) -g^{\rho \mu }\left(
R^{\nu }\psi \right) \right] ,  \label{eq24}
\end{equation}%
from both equations (\ref{eq23}) and (\ref{eq24}) we obtain the equation of
motion for the massless vector field $R^{\mu }\psi $
\begin{equation}
(\nabla _{\nu }+\tilde{\tau}_{\nu })(\nabla _{\rho }+\tilde{\tau}_{\rho
}\gamma )\left[ g^{\rho \nu }\left( R^{\mu }\psi \right) -g^{\rho \mu
}\left( R^{\nu }\psi \right) \right] =0\,,  \label{eqm18}
\end{equation}

We use a specific representation of the DKP algebra in which the singular $%
\gamma $ matrix is
\begin{equation*}
\gamma =\mbox{diag}(\lambda ,\lambda ,\lambda ,\lambda ,1-\lambda ,1-\lambda
,1-\lambda ,1-\lambda ,1-\lambda ,1-\lambda ,),
\end{equation*}%
from the condition $\gamma ^{2}-\gamma =0$ implies that $\lambda =0$ or $1$.
In Minkowski space--time, the value $\lambda =0$ reproduces the
electromagnetic field \cite{CQG}, thus, we restrict our attention for the $%
\lambda =0$ case only.\ Then in this representation the DKP field $\psi $ is
now a 10-component column vector
\begin{equation*}
\psi =\left(
\begin{array}{c}
\psi ^{0},\psi ^{1},\psi ^{2},\psi ^{3},\psi ^{23},\psi ^{31},\psi
^{12},\psi ^{10},\psi ^{20},\psi ^{30}%
\end{array}%
\right) ^{T}\;,
\end{equation*}%
where $\psi ^{a}$ ($a=0,1,2,3$) and $\psi ^{ab}$ behave, respectively, as a
4-vector and an antisymmetric tensor under \textit{Lorentz} transformations
on the Minkowski space. And we also get%
\begin{equation*}
R^{\mu }\psi =\left(
\begin{array}{c}
\psi ^{\mu },0,0,0,0,0,0,0,0,0%
\end{array}%
\right) ^{T}\,\ \ ,~\ \ ~R^{\mu \nu }\psi =\left(
\begin{array}{c}
\psi ^{\mu \nu },0,0,0,0,0,0,0,0,0%
\end{array}%
\right) ^{T}\,\
\end{equation*}%
\begin{equation*}
\gamma \psi =\left(
\begin{array}{c}
0,0,0,0,\psi ^{23},\psi ^{31},\psi ^{12},\psi ^{10},\psi ^{20},\psi ^{30}%
\end{array}%
\right) ^{T}\,\ \ ,~\ \ ~\,\ R^{\mu }\gamma \psi =0\ \ ,~\ \ ~\,\ R^{\mu \nu
}\gamma \psi =\left(
\begin{array}{c}
\psi ^{\mu \nu },0,0,0,0,0,0,0,0,0%
\end{array}%
\right) ^{T}\;.
\end{equation*}

Then, we get the following relations among $\psi $ components
\begin{equation}
i\,\psi _{\mu \nu }=\nabla _{\mu }\psi _{\nu }-\nabla _{\nu }\psi _{\mu }
\label{eq26}
\end{equation}%
which leads to the equation of motion for the spin $1$ sector of the
massless DKP field in Lyra space-time
\begin{equation}
(\nabla _{\mu }+\tilde{\tau}_{\mu })\left( {\nabla }^{\mu }\psi ^{\nu }-{%
\nabla }^{\nu }\psi ^{\mu }\right) =0\,.
\end{equation}%
In terms of these components the DKP\ action (\ref{lag-cv}) is written as
\begin{equation}
S_{1}=\int dx~\phi ^{4}e\left( \frac{i}{2}\psi ^{\ast \mu \nu }\left( \nabla
_{\mu }\psi _{\nu }-\nabla _{\nu }\psi _{\mu }\right) -\frac{i}{2}\psi ^{\mu
\nu }\left( \nabla _{\mu }\psi _{\nu }^{\ast }-\nabla _{\nu }\psi _{\mu
}^{\ast }\right) +\frac{1}{2}\psi ^{\ast \mu \nu }\psi _{\mu \nu }\right) \,.
\end{equation}

Then, using the equation (\ref{eq26}) in the action above we obtain the
action for the spin 1\ sector of the massless DKP field in Lyra manifold is
\begin{equation}
S_{1}=-\frac{1}{2}\int d^{4}x~\phi ^{4}e\left( \nabla _{\mu }\psi _{\nu
}^{\ast }-\nabla _{\nu }\psi _{\mu }^{\ast }\right) \left( \nabla ^{\mu
}\psi ^{\nu }-\nabla ^{\nu }\psi ^{\mu }\right)  \label{eq-pr29}
\end{equation}%
where $\nabla _{\mu }\psi _{\nu }-\nabla _{\nu }\psi _{\mu }$ is written as
\begin{equation}
\nabla _{\mu }\psi _{\nu }-\nabla _{\nu }\psi _{\mu }=\frac{1}{\phi }%
(\partial _{\mu }\psi _{\nu }-\partial _{\nu }\psi _{\mu })-\frac{1}{\phi }%
(\psi _{\mu }\partial _{\nu }-\psi _{\nu }\partial _{\mu })\ln \left( \frac{%
\phi }{\bar{\phi}}\right)
\end{equation}

\subsection{On the solution of the gauge condition for the vectorial field}

The gauge transformation (\ref{lyraEQ-38}) now reads
\begin{equation*}
\psi ^{\prime }{}^{\mu }=\psi ^{\mu }+\Phi ^{\mu }\quad ,\quad \psi ^{\prime
}{}^{\mu \nu }=\psi ^{\mu \nu }
\end{equation*}%
and the gauge condition for the massless DKP field is given by%
\begin{equation*}
{\nabla }_{\mu }\Phi _{\nu }-{\nabla }_{\nu }\Phi _{\mu }=0\,,
\end{equation*}%
or, explicitly,%
\begin{equation*}
\frac{1}{\phi }\left( \partial _{\mu }\Phi _{\nu }-\partial _{\nu }\Phi
_{\mu }\right) -2\Phi _{\rho }\tilde{\Gamma}_{\;\,[\mu \nu ]}^{\rho }=0
\end{equation*}%
Using%
\begin{equation}
2\tilde{\Gamma}_{\;\,[\mu \nu ]}^{\rho }=\tilde{\Gamma}_{\;\,\mu \nu }^{\rho
}-\tilde{\Gamma}_{\;\,\nu \mu }^{\rho }\,=\frac{1}{\phi }\left( \delta _{\mu
}^{\rho }\partial _{\nu }-\delta _{\nu }^{\rho }\partial _{\mu }\right) \ln
\left( \frac{\phi }{\bar{\phi}}\right)
\end{equation}%
we have
\begin{equation}
\partial _{\mu }\Phi _{\nu }-\partial _{\nu }\Phi _{\mu }-\frac{1}{\phi }%
\left( \Phi _{\mu }\partial _{\nu }-\Phi _{\nu }\partial _{\mu }\right) \ln
\left( \frac{\phi }{\bar{\phi}}\right) =0.  \label{eq-pr31}
\end{equation}

The gauge covariance of the system is assured since this equation has
solution. Let us begin looking for the simplest case when $\Phi_{\mu}
=\partial _{\mu }\Lambda $, then (\ref{eq-pr31}) becomes%
\begin{equation}
\left( \partial _{\mu }\Lambda \partial _{\nu }-\partial _{\nu }\Lambda
\partial _{\mu }\right) \ln \left( \frac{\phi }{\bar{\phi}}\right) =0,
\label{lambda-0}
\end{equation}%
therefore, if we choose
\begin{equation}
\Lambda \left( x\right) =\ln \left( \frac{\phi }{\bar{\phi}}\right) ,
\label{Lambda}
\end{equation}%
(\ref{lambda-0}) is satisfied.

Notwithstanding, this is not the most general solution. Another possible
solution is found by solving%
\begin{equation}
\partial _{\mu }\Phi _{\nu }-\Phi _{\nu }\partial _{\mu }\ln \left( \frac{%
\phi }{\bar{\phi}}\right) =0,  \label{L1}
\end{equation}%
or
\begin{equation}
\partial _{\mu }\Phi _{\nu }-\Lambda _{\mu }\Phi _{\nu
}=0\quad,\qquad\Lambda _{\mu }\equiv \partial _{\mu }\ln \left( \frac{\phi }{%
\bar{\phi}}\right)  \label{L2}
\end{equation}%
where $\Lambda _{\mu }$\ is considered as a prescribed function (part of the
external background). The solution of the equations (\ref{L1}) or (\ref{L2})
is
\begin{equation}
\Phi _{\nu }\left( x\right) =\Phi _{\nu }\left( x_{0}\right) \exp \left(
\int_{\gamma }d\gamma ^{\mu }\Lambda _{\mu }\right) =\Phi _{\nu }\left(
x_{0}\right) \left( \frac{\phi }{\bar{\phi}}\right)
\end{equation}%
where $\gamma $ is a particular trajectory joining the points $x$ and $x_{0}$%
, $\Phi _{\nu }\left( x_{0}\right) $ is a constant.

\bigskip A specialization to $\phi /\bar{\phi}\equiv cte$ gives%
\begin{equation}
\Phi _{\nu }\left( x\right) =\left( \frac{\phi \left( x\right) }{\bar{\phi}%
\left( x\right) }\right) \Phi _{\nu }\left( x_{0}\right) \equiv cte.
\end{equation}%
\qquad But, if $\phi ,\bar{\phi}\equiv cte$ then (\ref{eq-pr31}) is%
\begin{equation}
\partial _{\mu }\Phi _{\nu }-\partial _{\nu }\Phi _{\mu }=0
\end{equation}
which admits a solution of the form%
\begin{equation}
\Phi _{\nu }\left( x\right) =\partial _{\nu }\Lambda \left( x\right)
\end{equation}%
with $\Lambda $\ an arbitrary scalar function.

In another hand, if $\phi =\bar{\phi}$, (\ref{eq-pr31}) again simplify in%
\begin{equation}
\partial _{\mu }\Phi _{\nu }-\partial _{\nu }\Phi _{\mu }=0
\end{equation}%
and an analogous argument imply that%
\begin{equation}
\Phi _{\nu }\left( x\right) =\partial _{\nu }\Lambda \left( x\right)
\end{equation}%
and $\Lambda $ arbitrary. Thus, we show that in Lyra space--time the
coupling between electromagnetic field and torsion preserves the local $%
U\left( 1\right) $ gauge symmetry.

\section{Conclusions and perspectives}

We have show how to couple massless, scalar and vector, fields to Lyra
manifold by means of the DKP theory. Equations of motion, conservation laws
and gauge symmetry was obtained using Schwinger Action Principle. Projecting
to select the two distinct spin sectors, we find that the scalar field
couples to the trace of the torsion and the gauge symmetry of the vectorial
sector is maintained if one uses the scale function to perform the gauge
transformation. These are completely new and unexpected results, which do
not occurs in others geometries like the Riemann-Cartan one \cite{CQG}.

From the Lyra geometry we can obtain the Riemannian one when the scale
functions is set $\phi =\bar{\phi}\equiv 1$, obtaining from (\ref{3ra-3})
and (\ref{ConservSpin}) the conservation laws for the DKP field in the
torsionless geometry,
\begin{equation}
\nabla _{\mu }T_{\nu }{}^{\mu }+\frac{1}{2}S^{\mu ab}R_{\nu \mu ab}=0\,,
\label{c-l-1}
\end{equation}
\begin{equation}
\nabla _{\mu }S^{\mu ab}=T^{ab}-T^{ba}\,.  \label{c-l-2}
\end{equation}%
It is possible to show that in the spin $0$ sector the spin density tensor $%
S^{\mu ab}=0$, thus (\ref{c-l-1}) is reduced to $\nabla _{\mu }T_{\nu
}{}^{\mu }=0$ wile the second equation, (\ref{c-l-2}), gives that the
energy-momentum tensor is symmetric, the well-known results for the
Klein-Gordon-Fock field. Also, for the spin $1$ sector the conservation laws
equations (\ref{c-l-1}) and (\ref{c-l-2}) reproduce the electromagnetic case.

It is worthwhile to note that the structure of the conservation laws (\ref%
{3ra-3}), (\ref{ConservSpin}), the trace relation (\ref{trace-identity}) and
the trace symmetry (\ref{2da-3}) are independent of the matter fields, does
not matter the mass or spin content. Thus, for example, such equations have
the same form for the spin $1/2$ field \cite{dirac-lyra} or for the massive
DKP field \cite{massiveDKPLyra}.

The trace relation (\ref{trace-identity}) and the trace symmetry (\ref{2da-3}%
) in the scalar sector of DKP field give a traceless energy-momentum tensor.
This is just related to the scale invariance of the Lyra geometry, even in
the massive situation. The results here obtained seem to indicate that Lyra
geometry would be a useful tool to implement a kind of conformal symmetry
for massive fields. To make precise this reason, is necessary a deeper study
of the relation among conformal invariance and Lyra scale transformations.

The investigation of the back reaction of the fields (at least for spin $0$,
$1/2$ and $1$) upon the geometry would show us how a propagation equation
for the scale function can be obtained. These studies are currently under
discussion.

\begin{center}
\textbf{Acknowledgments}
\end{center}

This work is supported by FAPESP grants 01/12611-7 (RC), 01/12584-0 (CAMM)
and 02/00222-9 (BMP). BMP also thanks CNPq for partial support.


\end{document}